# A model of magneto-electric multipoles

S. W. Lovesey[1,2]

1. ISIS Facility, STFC, Oxfordshire OX11 0QX, UK

2. Diamond Light Source Ltd, Oxfordshire OX11 0DE, UK

**Abstract**

An ancient Hamiltonian of electrons with entangled spin and orbital degrees of freedom is re-examined as a model of magneto-electric multipoles. In the model, a magnetic charge and simple quantum rotator are tightly locked in action, some might say they are enslaved entities. It is shown that magneto-electric multipoles almost perfectly accord with those inferred from an analysis of magnetic neutron diffraction data on a ceramic superconductor (YBCO) in the pseudo-gap phase. Nigh on perfection between Stone's model and inferred magneto-electric multipoles is achieved by addition to the original model of a crystal-field potential appropriate for the magnetic space-group used in the published data analysis.

**1. Introduction**

In a quest to improve knowledge about magneto-electric multipoles (MEs) we re-visit a quantum model introduced by Stone [1, 2] of electrons with tightly locked spin and orbital degrees of freedom. We have in mind application to a high-$T_c$ superconductor (YBCO) that has been the subject of elegant investigations using diffraction of polarized neutrons [3, 4, 5]. Stone's original model addressed anomalies in quantum-field theory so it is little wonder that it does not immediately address MEs inferred from diffraction data on a ceramic superconductor [6]. To further the cause in materials physics, we consider additions to the original model that arise from an electrostatic crystal-field potential experienced by magnetic Cu ion in YBCO.

It has been known for some time that MEs contribute to resonant x-ray diffraction, and results from a number of experiments have been published, e.g., [7, 8, 9, 10]. More recently, it has been shown that MEs contribute in magnetic neutron scattering [11]. An additional, independent development is simulations of electronic structure that estimate MEs [12].

The communication starts with a resumé of Stone's original model, which is followed by an enumeration of some of its MEs. Results quoted are derived using techniques that are standard in quantum mechanics. (An Appendix surveys tools employed in calculations, for the convenience of newcomers.) MEs in Stone's model are an almost perfect match to those inferred from neutron diffraction data gathered on YBCO in the pseudo-gap phase. The ME that prevents the perfect match, between model MEs and inferred MEs, is attributed in Section 4 to a component of the electrostatic crystal-field potential experienced by a magnetic Cu ion in YBCO. This discourse is followed by a discussion of findings in Section 5.

**2. Stone's model**

An operator is labelled magneto-electric if it is parity-odd and time-odd, with discrete time-signature $\sigma_\theta = -1$ and parity-signature $\sigma_\pi = -1$. Perhaps best known is a spin anapole discussed by Zel'dovich in an investigation of electromagnetic interactions that violate parity [13], and the panoply of MEs reviewed by Dubovik and Tugushev [14] includes the Majorana particle.

Stone introduced and analysed a Hamiltonian built with a scalar magneto-electric operator, or magnetic monopole, (**S·n**) coupled to a rotator defined by a Hamiltonian $\mathbf{L}^2$. Here, **S**, **n** and **L** are operators for electron spin, electric-dipole moment and orbital angular momentum, with $[L_\alpha, n_\beta] = i\varepsilon_{\alpha\beta\gamma} n_\gamma$. The Hamiltonian in question is,

$$\mathcal{H} = \mathbf{L}^2 - \lambda(\mathbf{S}\cdot\mathbf{n}), \tag{2.1}$$

and we will consider its properties in the limit of a strong coupling when $\lambda \to \infty$. Both operators in $\mathcal{H}$ obey $\sigma_\theta \sigma_\pi = +1$. (Our notation in (2.1) differs from Stone in that we set $2I = 1$ and $\mu = \lambda/2$.)

Given explicit spin and orbital variables in $\mathcal{H}$ its eigenstates are best constructed using j-j coupling. Basis states are,

$$|s l jm\rangle = \sum_{\sigma\mu} |s\sigma\rangle |l\mu\rangle (s\sigma l\mu|jm), \tag{2.2}$$

in which there are $(2j + 1)$ projections m, spin $s = \frac{1}{2}$ and orbital angular momentum $l = j \pm s$. The standard Clebsch-Gordan coupling coefficient $(s\sigma l\mu|jm)$ is purely real and it is related to a Wigner 3j-symbol [15, 16],

$$(a\alpha b\beta|IM) = (-1)^{-a+b-M}\sqrt{(2I+1)} \begin{pmatrix} a & b & I \\ \alpha & \beta & -M \end{pmatrix}. \tag{2.3}$$

States $|s l jm\rangle$ are orthogonal and $\langle s l jm|s l' j'm'\rangle = \delta_{m,m'} \delta_{l,l'} \delta_{j,j'}$.

Turning to operators in $\mathcal{H}$, $\langle s l jm|\mathbf{L}^2|s l' j'm'\rangle = \delta_{m,m'} \delta_{l,l'} \delta_{j,j'} l(l+1)$ and,

$$\langle s l jm|(\mathbf{S}\cdot\mathbf{n})|s l' j'm'\rangle = (1/2)\, \delta_{m,m'} \delta_{j,j'}, \tag{2.4}$$

with $l - l' = \pm 1$. The sign on the right-hand side of (2.4) is correct for s-$l$ coupling that is in use in our communication. From these two matrix elements, for $\mathbf{L}^2$ and (**S·n**), it follows that matrix elements of $\mathcal{H}$ are diagonal with respect to total angular momentum j and its projections m. States of $\mathcal{H}$ form a (2 x 2) matrix with elements labelled by $l_+ = j + (\frac{1}{2})$ and $l_- = j - (\frac{1}{2})$. Using $(a + b) = l_+(l_+ + 1)$ and $(a - b) = l_-(l_- + 1)$ the ground-state energy is,

$$E = a - c, \tag{2.5}$$

with $c = \sqrt{[b^2 + (\lambda/2)^2]}$ and eigenstates,

$$[|s l_+ jm\rangle + \alpha|s l_- jm\rangle]/\sqrt{(1 + \alpha^2)}, \text{ with } \alpha = (\lambda/2)/(c - b). \tag{2.6}$$

In the limit of strong coupling $\alpha \to 1$, and we denote the corresponding ground-state by,

$$|\varphi_m\rangle = [|s l_+ jm\rangle + |s l_- jm\rangle]/\sqrt{2}. \tag{2.7}$$

Expectation values of operators with respect to $|\varphi_m\rangle$ are constructed using standard tools in quantum mechanics that are surveyed in an Appendix [15, 16].

**3. Multipoles**

Operators that participate in the diffraction of radiation include a magnetic monopole (**S·n**), magnetic moment $\boldsymbol{\mu} = (\mathbf{L} + 2\mathbf{S})$, and an electric dipole **n**. An alternative to the tools for expectation values in the Appendix is use of operator equivalents. Thus, $\boldsymbol{\mu} \propto \mathbf{J}$ where $\mathbf{J}$ is the operator for total angular momentum, and the coefficient of proportionality is the Landé factor. We find,

$$\langle \varphi_m|(\mathbf{S}\cdot\mathbf{n})|\varphi_m\rangle = 1/2, \quad \langle \varphi_m|\boldsymbol{\mu}_q|\varphi_m\rangle = \delta_{q,0}\, m\, [1 + \{4j(j+1)\}^{-1}], \tag{3.1}$$

$$\langle \varphi_m|\mathbf{n}_q|\varphi_m\rangle = \delta_{q,0}\, [m/\{2j(j+1)\}],$$

where q denotes a spherical component and $q = 0 \equiv z$. Evidently, expectation values of the magnetic moment and electric dipole moment are different from zero and can be positive or negative. In contrast to this finding, spin and orbital anapoles are zero;

$$\langle \varphi_m|(\mathbf{S} \times \mathbf{n})|\varphi_m\rangle = \langle \varphi_m|\boldsymbol{\Omega}|\varphi_m\rangle = 0. \tag{3.2}$$

Here, $\boldsymbol{\Omega} = (\mathbf{L} \times \mathbf{n}) - (\mathbf{n} \times \mathbf{L}) = i[\mathbf{L}^2, \mathbf{n}]$ with $[L_\alpha, \Omega_\beta] = i\, \varepsilon_{\alpha\beta\gamma}\Omega_\gamma$ and $[n_\alpha, \Omega_\beta] = 2i\, (\delta_{\alpha\beta} - n_\alpha n_\beta)$ [2, 7]. Note that $\mathbf{L}$ is orthogonal to both $\mathbf{n}$ and $\boldsymbol{\Omega}$, but $\mathbf{n}$ and $\boldsymbol{\Omega}$ satisfy $\boldsymbol{\Omega}\cdot\mathbf{n} + \mathbf{n}\cdot\boldsymbol{\Omega} = 0$ [7]. The result $\mathbf{L}\cdot\mathbf{n} = 0$ accounts for the absence of orbital angular momentum as a magnetic monopole.

Of the five charge quadrupoles $(q = 0, \pm 1, \pm 2)$ only the diagonal $q = 0$ is different from zero, namely,

$$\langle \varphi_m|(3S_z n_z - \mathbf{S}\cdot\mathbf{n})|\varphi_m\rangle = \{4j(j+1)\}^{-1}[j(j+1) - 3m^2]. \tag{3.3}$$

Degeneracy with respect to m in expressions (3.1) and (3.3) will be removed by an exchange field and a thermal average $\langle\, ...\, \rangle$. Two cases $m \to \langle J_z \rangle$ and $m^2 \to \langle (J_z)^2 \rangle$ for various j using the molecular-field approximation are found in reference [17]. In general, values of $\langle J_z \rangle$ and $\langle (J_z)^2 \rangle$ are largely independent, and $\langle J_z \rangle$ can be very small or zero.

## 4. Magnetic charge in YBCO

Neutron Bragg diffraction data gathered on YBCO in the pseudo-gap phase have been successfully interpreted using sites 4i for Cu ions in the magnetic space-group C2/m' (#12.61) [5, 6]. With respect to the parent P4/mmm-type structure, with Cu ions in sites 2g (symmetry 4mmm), bases vectors (x, y, z) for C2/m' are $\{(1, -1, 0), (1, 1, 0), (0, 0, 1)\}$ and sites 4i have symmetry m'.

The magnetic monopole, or charge, $(\mathbf{S}\cdot\mathbf{n})$ is not visible in neutron diffraction [11]. Experimental data are consistent with null values of spin and orbital anapoles, in-plane (x-y) magnetic moments and the charge quadrupole with projection $\pm 1$. The charge quadrupole $q = \pm 2$, however, appears to be different from zero. Indeed, charge quadrupoles with $q = 0$ and $q = \pm 2$ alone appear to provide a successful interpretation of the available data. Casting an eye back to previous findings for Stone's model, we have a charge quadrupole $q = 0$ in (3.3) together with other results that accord with inferred MEs, which leaves the absence of a charge quadrupole with $q = \pm 2$ a shortfall in an otherwise nice story.

The desired quadrupole, $q = \pm 2$, is missing in Stone's model because it possesses cylindrical symmetry. This high symmetry can be reduced to a desired level by addition of an electrostatic crystal-field potential. Calculations of crystal-field potentials are notoriously unreliable, and even relative signs of contributions can be at fault. In consequence, it is usual practice to exploit site symmetry to fix a grand structure of a crystal-field potential and determine phenomenological parameters therein by fits to relevant data [18, 19]. In the present case, Cu ions in C2/m' have site symmetry m, also denoted $C_{1h}$ or $C_s$, which requires the potential to be unchanged by the transformation (x, y, z) → (x, −y, z). The lowest-order contribution with q = ± 2 to the potential that is allowed takes the form $O^2_2 = [Y^2_{+2} + Y^2_{-2}] \propto (x^2 - y^2)$, where $Y^k_q$ is a standard spherical harmonic of rank k and projection q [15]. (A contribution to the potential with q = ± 1 would induce a quadrupole with like projections and these are absent from inferred MEs.)

The new Hamiltonian is $\{\mathcal{H} + gO^2_2\}$, where the coupling constant g has both unknown magnitude and sign. The secular equation in a basis with projections m and (m + 2) has non-trivial solutions for,

$$(E - a + \lambda/2)^2 = \{g\langle\varphi_{m+2}|O^2_2|\varphi_m\rangle\}^2. \qquad (4.1)$$

The corresponding ground-state wavefunction is,

$$|\psi\rangle = [|\varphi_m\rangle + \eta|\varphi_{m+2}\rangle]/\sqrt{2}, \qquad (4.2)$$

where η = ± 1, with η = + 1 (− 1) when $[g\langle\varphi_{m+2}|O^2_2|\varphi_m\rangle]$ is negative (positive).

Anapoles are not changed by addition of the crystal-field potential, and remain in accord with the inferred MEs. For charge quadrupoles q = 0 and q = + 2 that are all-important in the interpretation of neutron diffraction data on YBCO,

$$\langle\psi|(3S_z n_z - \mathbf{S}\cdot\mathbf{n})|\psi\rangle = \{4j(j + 1)\}^{-1}[j(j + 1) - 3(m^2 + 2m + 2)], \qquad (4.3)$$

$$\langle\psi|(S_x n_x - S_y n_y)|\psi\rangle = -\eta\{8j(j + 1)\}^{-1}[(j - m - 1)(j - m)(j + m + 1)(j + m + 2)]^{1/2}.$$

Recall that x- and y-axes are in a plane normal to the c-axis of the parent P4/mmm-type structure, and they subtend an angle 45° with respect to a- and b-axes.

## 5. Discussion

Stone's model [1] of a magnetic charge coupled to a quantum rotator supports many magneto-electric multipoles visible in the diffraction of neutrons and x-rays by a magnetic material. Absent from the model are the spin anapole (**S x n**) and its orbital analogue, whereas the expectation value of the electric dipole **n** can be different from zero. We provide tools that enable calculation of all multipoles in the model.

When the allowed magneto-electric multipoles are confronted with those for a ceramic superconductor YBCO, inferred from neutron Bragg diffraction data using a monoclinic magnetic space-group C2/m' [5, 6], Stone's original model falls short because of its cylindrical symmetry. We demonstrate that addition of the electrostatic crystal-field potential acting on a magnetic Cu ion reduces the spatial symmetry in an appropriate manner, and a model that includes the potential accords with magneto-electric multipoles inferred from diffraction data. The story is perfect when the magnetic and electric dipole moments have expectation values that are too small to contribute in the analysis of available experimental data.

**Acknowledgement**

We are grateful to Professor E Balcar for comments on an early draft of the communication.

**Appendix**

A spherical tensor-operator $U^K_Q$ has rank K and (2K + 1) projections Q, with $-K \leq Q \leq K$. A matrix element of $U^K_Q$ obeys the Wigner-Eckart Theorem [15],

$$\langle s l j m | U^K_Q | s l' j' m' \rangle = (-1)^{j-m} (s l j \| U^K \| s l' j') \begin{pmatrix} j & K & j' \\ -m & Q & m' \end{pmatrix}, \tag{A.1}$$

in which $(slj\|U^K\|sl'j')$ is a reduced matrix-element (RME). The phase of the RME is intimately related to the discrete symmetries of $\mathbf{U}^K$, as we see in two identities (A.2). Hereafter, quantum labels are abbreviation by $\theta = slj$ and $\theta' = sl'j'$.

An operator is Hermitian if it satisfies $(U^K_Q)^\dagger = (-1)^Q U^K_{-Q}$. For such an operator the complex conjugate of the corresponding RME satisfies two fundamental identities [16],

$$(\theta\|U^K\|\theta')^* = (-1)^{j-j'} (\theta'\|U^K\|\theta) = \sigma_\theta \sigma_\pi (-1)^K (\theta\|U^K\|\theta'), \tag{A.2}$$

where $\sigma_\theta$ and $\sigma_\pi$ are the time and parity signatures of $\mathbf{U}^K$, respectively. The first identity gives $(\theta\|\mathbf{\Omega}\|\theta') = -(\theta'\|\mathbf{\Omega}\|\theta)$ and $(\theta\|(\mathbf{S} \times \mathbf{n})\|\theta') = -(\theta'\|(\mathbf{S} \times \mathbf{n})\|\theta)$ for $j = j'$, and the two relations account for results in (3.2). Identities in (A.2) apply also to operators acting solely on orbital degrees of freedom. As examples we cite RMEs for Hermitian operators $\mathbf{\Omega}$ and $\mathbf{n}$,

$$(l\|\mathbf{\Omega}\|l') = i \, [l(l+1) - l'(l'+1)] \, (l\|\mathbf{n}\|l'), \tag{A.3}$$

that is purely imaginary and symmetric with respect to an interchange of $l$ and $l'$ ($i = \sqrt{(-1)}$), which are exact opposites of,

$$(l\|\mathbf{n}\|l') = (-1)^l \sqrt{[(2l+1)(2l'+1)]} \begin{pmatrix} l & 1 & l' \\ 0 & 0 & 0 \end{pmatrix}. \tag{A.4}$$

The RME (A.4) vanishes unless $l + l'$ is odd, because $\mathbf{n}$ is parity-odd. One finds $(l\|\mathbf{n}\|l-1) = \sqrt{l}$.

A tensor product $X^K_Q$ of two commuting Hermitian operators, $\mathbf{z}^a$ and $\mathbf{y}^b$, is defined by analogy with (2.2). A complex phase is introduced to make $X^K_Q$ Hermitian. Let $\mathbf{z}^a$ act on spin and $\mathbf{y}^b$ act on orbital variables. We define,

$$X^K_Q = (-i)^{a+b+K} \sum_{\alpha\beta} z^a_\alpha y^b_\beta \, (a\alpha b\beta | KQ), \tag{A.5}$$

and the associated RME is,

$$(\theta\|X^K\|\theta') = (-i)^{a+b+K} (s\|z^a\|s) (l\|y^b\|l') \, W^{(a,b)K}(\theta, \theta'), \tag{A.6}$$

with a unit tensor,

$$W^{(a,b)K}(\theta, \theta') = [(2j+1)(2K+1)(2j'+1)]^{1/2} \begin{Bmatrix} \tfrac{1}{2} & \tfrac{1}{2} & a \\ l & l' & b \\ j & j' & K \end{Bmatrix}. \tag{A.7}$$

The magnitude of the 9j-symbol in (A.7) is unchanged by an even or odd exchange of columns or rows, but an odd exchange changes its sign by a factor $(-1)^\Re$ with $\Re = 1 + a + l + l' + b + j + j' + K$ [15, 16]. Since $s = \tfrac{1}{2}$, the rank of $\mathbf{z}^a$ has two values $a = 0$ or 1; $(s\|z^0\|s) = \sqrt{2}$ and $(s\|z^1\|s) = \sqrt{(3/2)}$.

Matrix elements of all operators in the main text are obtained using (A.1) and (A.5) together with appropriate RMEs evaluated with j = j'. For example, $\langle\theta m|\mathbf{n}_q|\theta'm'\rangle$ may be obtained using a = 0, b = 1 and $y^1_q = \mathbf{n}_q$ together with (A.4). Similarly, $\langle\theta m|\mathbf{\Omega}_q|\theta'm'\rangle$ is obtained with a = 0, b = 1 and the RME (A.3). Multipoles for magnetic charge possess a = 1 and $y^1_q = \mathbf{n}_q$. Matrix elements of the magnetic monopole and anapole are derived using $X^0_0 = (1/\sqrt{3})\,(\mathbf{S}\cdot\mathbf{n})$ and $X^1_q = -(1/\sqrt{2})\,(\mathbf{S}\times\mathbf{n})_q$. Quadrupoles for magnetic charge that are of interest in Section 4 are derived from $X^2_q$ with q = 0 and ± 2, e.g., $X^2_{+2} = (\mathbf{S}_{+1}\,\mathbf{n}_{+1})$ that is converted to Cartesian form using the relation $\mathbf{n}_{+1} = -(1/\sqrt{2})\,(\mathbf{n}_x + i\mathbf{n}_y)$ and a similar relation for $\mathbf{S}_{+1}$.

Analytic expressions for 3j- and 9j-symbols required in results reported in the main text are found in reference [15], for example, apart from a 9j-symbol in $W^{(1,1)2}(\theta, \theta')$ for which we find the result,

$$\begin{Bmatrix} \tfrac{1}{2} & \tfrac{1}{2} & 1 \\ l_+ & l_- & 1 \\ j & j & 2 \end{Bmatrix} = -(1/6)\{(2j+1)\}^{-1}[(2j-1)(2j+3)/\{10j(j+1)\}]^{1/2}, \quad (A.8)$$

where $l_+ = j + (\tfrac{1}{2})$ and $l_- = j - (\tfrac{1}{2})$.